\documentclass[%
 reprint, amsfonts,amssymb,amsmath,longbibliography,superscriptaddress]{revtex4-2}

\usepackage{graphicx}
\usepackage{dcolumn}
\usepackage{bm}
\usepackage[colorlinks=true,citecolor=blue,linkcolor=blue,urlcolor=blue]{hyperref}
\usepackage{color}
\usepackage{xcolor}
\usepackage[normalem]{ulem}

\usepackage{amsmath}
\usepackage{amssymb}

\begin{document}

\title{Multiorbital character of the density wave in trilayer nickelate superconductors}

\author{Abhi Suthar}
\affiliation{Max-Planck-Institute for Solid State Research, Heisenbergstra{\ss}e 1, 70569 Stuttgart, Germany}

\author{Vignesh Sundaramurthy}
\affiliation{Max-Planck-Institute for Solid State Research, Heisenbergstra{\ss}e 1, 70569 Stuttgart, Germany}

\author{Matías Bejas}
\affiliation{Facultad de Ciencias Exactas, Ingeniería y Agrimensura and Instituto de Física Rosario (UNR-CONICET),
Avenida Pellegrini 250, 2000 Rosario, Argentina}

\author{Congcong Le}
\affiliation{RIKEN Interdisciplinary Theoretical and Mathematical Sciences (iTHEMS), Wako, Saitama 351-0198, Japan}

\author{Pascal Puphal}
\affiliation{Max-Planck-Institute for Solid State Research, Heisenbergstra{\ss}e 1, 70569 Stuttgart, Germany}

\author{Pablo Sosa-Lizama}
\affiliation{Max-Planck-Institute for Solid State Research, Heisenbergstra{\ss}e 1, 70569 Stuttgart, Germany}

\author{Armin Schulz}
\affiliation{Max-Planck-Institute for Solid State Research, Heisenbergstra{\ss}e 1, 70569 Stuttgart, Germany}

\author{Jürgen Nuss}
\affiliation{Max-Planck-Institute for Solid State Research, Heisenbergstra{\ss}e 1, 70569 Stuttgart, Germany}

\author{Masahiko Isobe}
\affiliation{Max-Planck-Institute for Solid State Research, Heisenbergstra{\ss}e 1, 70569 Stuttgart, Germany}

\author{Peter A. van Aken}
\affiliation{Max-Planck-Institute for Solid State Research, Heisenbergstra{\ss}e 1, 70569 Stuttgart, Germany}

\author{Y. Eren Suyolcu}
\affiliation{Max-Planck-Institute for Solid State Research, Heisenbergstra{\ss}e 1, 70569 Stuttgart, Germany}

\author{Matteo Minola}
\affiliation{Max-Planck-Institute for Solid State Research, Heisenbergstra{\ss}e 1, 70569 Stuttgart, Germany}

\author{Andreas P. Schnyder}
\affiliation{Max-Planck-Institute for Solid State Research, Heisenbergstra{\ss}e 1, 70569 Stuttgart, Germany}

\author{Xianxin Wu}
\affiliation{Institute of Theoretical Physics, Chinese Academy of Sciences, Beijing 100190, China}

\author{Bernhard Keimer}
\affiliation{Max-Planck-Institute for Solid State Research, Heisenbergstra{\ss}e 1, 70569 Stuttgart, Germany}

\author{Giniyat Khaliullin}
\affiliation{Max-Planck-Institute for Solid State Research, Heisenbergstra{\ss}e 1, 70569 Stuttgart, Germany}

\author{Andrés Greco}
\email[]{agreco@fceia.unr.edu.ar}
\affiliation{Facultad de Ciencias Exactas, Ingeniería y Agrimensura and Instituto de Física Rosario (UNR-CONICET),
Avenida Pellegrini 250, 2000 Rosario, Argentina}

\author{Matthias Hepting}
\email[]{hepting@fkf.mpg.de}
\affiliation{Max-Planck-Institute for Solid State Research, Heisenbergstra{\ss}e 1, 70569 Stuttgart, Germany}

\date{\today}

\begin{abstract}
Ruddlesden-Popper nickelates exhibit high-temperature superconductivity closely intertwined with charge and spin density waves. However, fundamental questions persist regarding the interplay between the associated density wave (DW) fluctuations and superconductivity, as well as the orbital character and symmetry underlying the DW instabilities. Here we utilize polarized Raman scattering to investigate the phononic and electronic Raman responses of the trilayer nickelate La$_4$Ni$_3$O$_{10}$ across its concomitant charge and spin density wave transitions. In addition to distinct phonon anomalies occurring below the transition temperature, we observe a depletion of continuum spectral weight up to 114\,meV and a pronounced peak centered at this energy. By combining momentum-selective information from polarized electronic Raman scattering with Raman-response model calculations based on a multiorbital Raman vertex in a reconstructed two-orbital DW state involving both Ni-3$d_{x^2 - y^2}$ and Ni-3$d_{z^2}$ orbitals, we identify 114\,meV as the energy scale $2\Delta_\mathrm{DW}$ of the DW gap, characterized by incoherent opening and non-mean-field behavior. Furthermore, the model calculations reveal that the corresponding $2\Delta_\mathrm{DW}$ peak has a multiorbital origin, requiring both orbital contributions and their mixing beyond single-orbital projections, thus shedding light on the nature of the DW instabilities in La$_4$Ni$_3$O$_{10}$. 
\end{abstract}

\maketitle


\section{Introduction}
\label{sec:intro}

Unconventional superconductors, including cuprates, iron pnictides, and heavy-fermion materials, host symmetry-broken phases and electronic instabilities that either compete with or facilitate superconductivity~\cite{Scalapino2012,Lee_review2006,Fernandes2022,Gegenwart2008,Bruin2013,Timusk1999}. Raman spectroscopy has provided key insights into these intertwined phases through phonon anomalies, collective electronic modes, spectral-weight redistribution, and low-frequency scaling behavior of the electronic continuum~\cite{Devereaux2007,Opel2000,Bakr2009,Gallais2013,Kretzschmar2016,loret2019}.

Recently, Ni-based materials, exhibiting either the infinite-layer or the La$_{n+1}$Ni$_n$O$_{3n+1}$ Ruddlesden–Popper (RP) crystal structure, have emerged as a distinct class of unconventional superconductors \cite{Nomura2022,Wang2024b}. For RP nickelates, superconducting transition temperatures ($T_c$) comparable to those of several cuprate high-temperature superconductors have been reported: The bilayer compound La$_3$Ni$_2$O$_7$ exhibits superconductivity above 80~K under pressures exceeding 14 GPa \cite{Sun2023}, while trilayer La$_4$Ni$_3$O$_{10}$ also becomes superconducting under pressure, albeit with lower $T_c$ values between 15 to 40 K~\cite{Zhu2024,Sakakibara2024,Li_2024HTS4310,Zhang2025}.

At ambient pressure, both La$_3$Ni$_2$O$_7$ and La$_4$Ni$_3$O$_{10}$ display charge density wave (CDW) order~\cite{zhang2020intertwined,Kakoi2024,Li2025STM,Khasanov2025,Fukamachi2001}, which is gradually suppressed under pressure before superconductivity emerges \cite{Sun2023,Zhu2024}, suggesting a competitive interplay between these phases. In La$_4$Ni$_3$O$_{10}$, a spin density wave (SDW) transition is concomitantly suppressed with the CDW under pressure~\cite{Khasanov2025}, while residual spin fluctuations that persist in the absence of long-range order might play an essential role for superconductivity, analogous to the spin-fluctuation mediated pairing in other unconventional superconductors~\cite{Scalapino2012}. 
In La$_3$Ni$_2$O$_7$, however, moderate pressure enhances the SDW transition~\cite{Khasanov2025a}, leading to a divergence of the CDW and SDW transition temperatures prior to the onset of superconductivity, suggesting a more complex interplay of electronic phases. More fundamentally, the basic properties of the density-wave (DW) instabilities in La$_3$Ni$_2$O$_7$ and La$_4$Ni$_3$O$_{10}$ are still under debate---including the size of the DW gaps, their momentum dependence, and the orbital character of the involved electronic states---further complicating efforts to determine how the DW orders intertwine with superconductivity.

For La$_4$Ni$_3$O$_{10}$, band structure calculations predict a multiorbital Fermi surface (FS) comprising dispersive bands with mixed Ni-3$d_{x^2 - y^2}$ and 3$d_{z^2}$ orbital character 
along with a flat band of primarily $d_{z^2}$ character ($\gamma$ band)
~\cite{Chen2024Trilayer,Yang2024,Zhang2024b,Huang2024,LaBollita2024b,Leonov2024,Tian2024,Zhang2024a,Zhang2025b}. Angle-resolved photoemission spectroscopy (ARPES) studies observed a comparable FS topology but reached different conclusions regarding the involvement and position of the $d_{z^2}$-dominated $\gamma$ band as well as the size of the DW gap~\cite{Li2017,du2024}. Infrared and ultrafast optical spectroscopy as well as scanning tunneling microscopy (STM) studies also reported conflicting DW gap magnitudes~\cite{Li2025SDW,Li2025STM}, with values as large as $2\Delta_\mathrm{DW} \approx 122$\,meV~\cite{Xu2025,Xu2025Pressure}. 

A Raman spectroscopy study extracted a gap value of $2\Delta_\mathrm{DW} \approx 110$\,meV and proposed that Ni $d_{x^2 - y^2}$ states drive the DW order, reminiscent of the dominant role of Cu $d_{x^2-y^2}$ orbitals in the low-energy physics of cuprates~\cite{Gim2025}. This orbital-selective assessment in Ref.~\cite{Gim2025} is based on identifying FS regions with large electronic band curvature and associating them with selection rules observed in the electronic Raman response. Along similar lines, studies have overlaid Raman-vertex form factors on the FS of RP nickelates and inferred from the distribution of form-factor weight which FS sheets and orbital states participate in DW formation~\cite{Shu2026,He2026}. However, while such Raman-vertex form factors obtained within the effective-mass approximation and plotted on the FS of effective single-band systems, such as cuprates, can provide useful guidance for identifying the electronic states that contribute to the Raman response~\cite{Devereaux2007}, this interpretation can be less straightforward in multiorbital systems. Indeed, recent work on Sr$_2$RuO$_4$, a material with several FS sheets and active orbitals, has shown that the Raman vertex must be treated as a matrix in orbital space and retained explicitly in the Raman-response calculation to identify the contributing electronic states~\cite{Blesio2024}. This motivates a multiorbital analysis of the Raman response of La$_4$Ni$_3$O$_{10}$, especially in view of a growing body of work emphasizing that key electronic instabilities and fluctuations in RP nickelates involve $d_{z^2}$ or mixed $d_{x^2 - y^2}$/$d_{z^2}$ states~\cite{Chen2024Trilayer,Yang2024,Zhang2024b,LaBollita2024b,Zhang2024a,Zhang2025b,Lu2025,Zhang2025d,gu23,Luo2023,le2025,Lechermann2023,Cao2024,Ryee2025}. 

In this work, we use polarization-resolved Raman scattering to investigate DW order in high-quality La$_4$Ni$_3$O$_{10}$ single crystals. In the phonon sector, we observe several spectral anomalies that carry fingerprints of the DW transition. In particular, high-energy oxygen bond-stretching modes, identified via density functional theory (DFT) based phonon calculations, show clear intensity changes across $T_\mathrm{DW}$, whereas frequency and linewidth renormalizations are subtle. Symmetric phonon lineshapes indicate modest electron-phonon coupling, suggesting that DW order is primarily driven by electronic instabilities of the FS~\cite{Jia2025} rather than electron-lattice interactions. In electronic Raman scattering, we detect spectral weight depletion with decreasing temperature, indicative of a gap opening with $2\Delta_\mathrm{DW} \approx 114$\,meV, broadly consistent with previous Raman~\cite{Gim2025} and optical conductivity results on La$_4$Ni$_3$O$_{10}$~\cite{Xu2025,Xu2025Pressure}. In addition, we observe a distinct polarization-dependent peak centered at the same energy. Using a Raman-response calculation that retains the multiorbital Raman vertex in a reconstructed two-orbital DW state, we identify this feature as the $2\Delta_\mathrm{DW}$ peak and show that it cannot be reproduced by single-orbital projections. Instead, the spectra require both Ni-$d_{x^2-y^2}$ and $d_{z^2}$ orbital contributions together with their mixing. We further unveil a non-BCS-like evolution of the gap opening and short-ranged DW correlations persisting above $T_\mathrm{DW}$, and discuss implications of the multiorbital DW fluctuations for superconductivity.

\begin{figure*}
    \centering
 \includegraphics[width=0.9\linewidth]{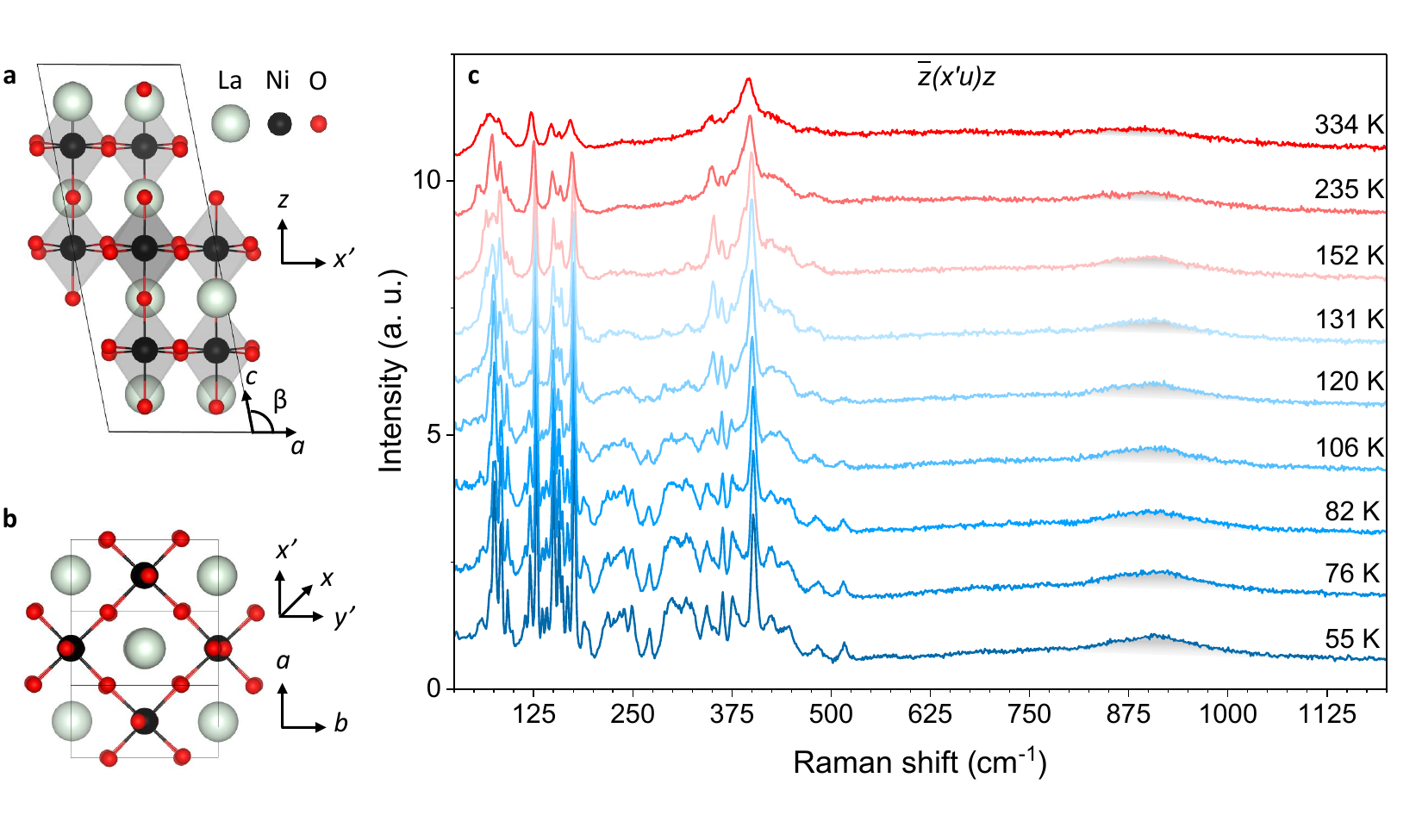}
    \caption [Temperature evolution of unpolarized Raman spectra in La$_4$Ni$_3$O$_{10}$]{Crystal structure and temperature dependence of Raman scattering in La$_4$Ni$_3$O$_{10}$. (a) Side view of the $P2_1/a$ unit cell of La$_4$Ni$_3$O$_{10}$, with the monoclinic angle $\beta$ between the $a$ and $c$ axes indicated. The $x'$ and $z$ directions of the laboratory frame of the Raman experiment are also indicated. 
    (b) Top view of the unit cell, with the $a$ and $b$ axes indicated, along with the $x'$, $x$, and $y'$ directions of the laboratory frame. (c) Raman spectra at various temperatures, acquired in $\bar z(x'u)z$ scattering configuration (without analyzer). Spectra above (below) the DW transition are colored in red (blue). All spectra are divided by the Bose-Einstein factor and offset in vertical direction by 1.25~a.u.\ for clarity.
    }
    \label{fig:1}
\end{figure*}

\section{Results}
\label{sec:results}

\begin{figure*}
    \centering
    \includegraphics[width=1\linewidth]{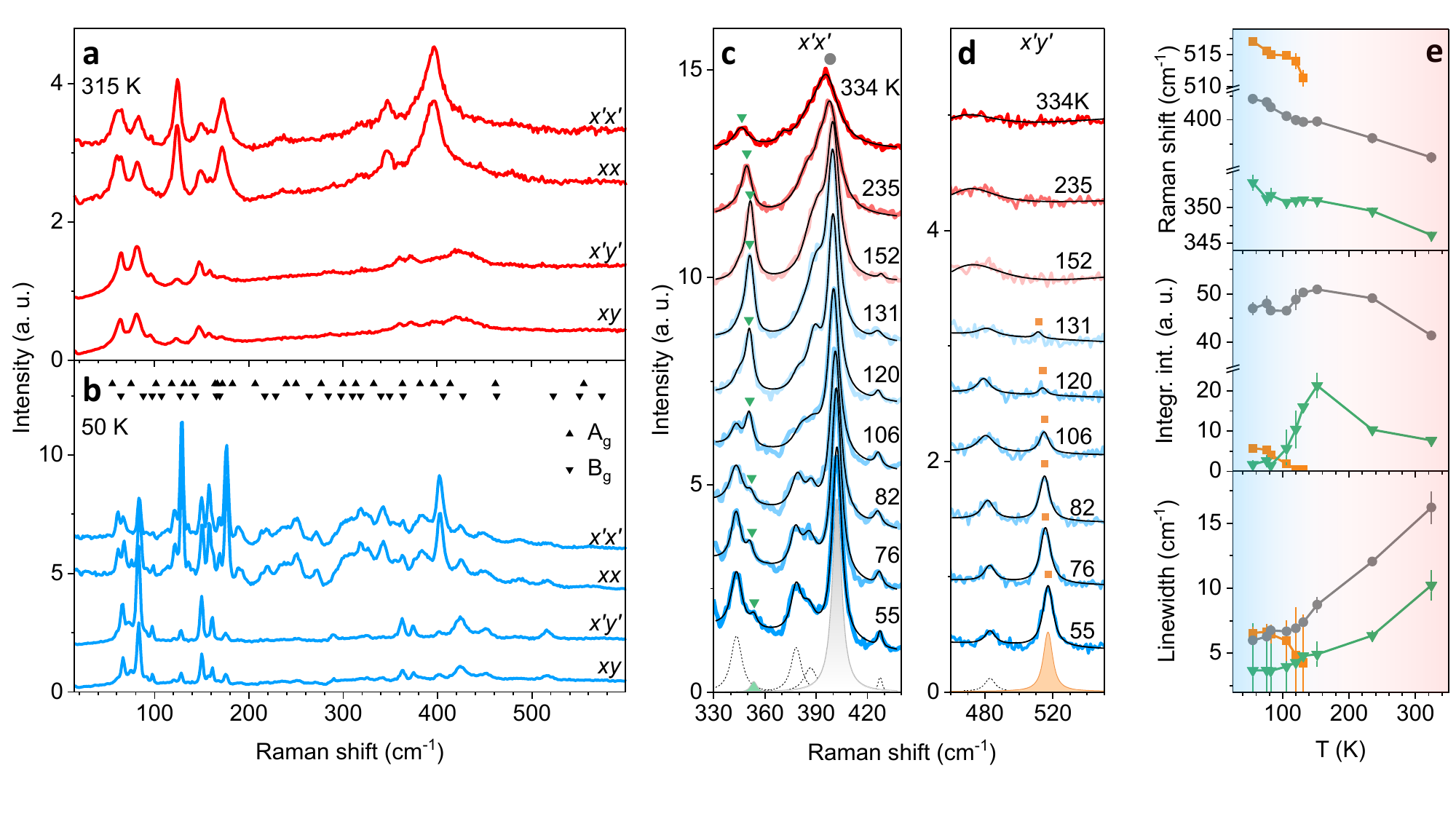}
    \caption{Phononic Raman scattering in La$_4$Ni$_3$O$_{10}$. (a, b) Polarization-resolved Raman spectra at 315~K (a) and 50 K (b), acquired in $x'x'$, $xx$, $x'y'$, and $xy$ scattering configuration, respectively. Black triangles in (b) indicate computed phonon frequencies in $A_g$ and $B_g$ symmetry, respectively. (c) Temperature dependence of selected phonon modes in the $x'x'$ channel. Solid black lines are fits of the data using Voigt profiles. The individual profiles of the 55 K fit are shown as dashed lines at the bottom of the panel. The profiles of phonon modes around 355 and 400\,cm$^{-1}$ are highlighted in green and gray, respectively. Green triangles and gray circles indicate the position of the modes at each temperature. (d) Temperature dependence of selected phonon modes in the $x'y'$ channel. The profile of a phonon mode around 515\,cm$^{-1}$ is highlighted in orange, and the orange squares indicate the position of this mode at several temperatures. Spectra in (a)-(d) are offset in vertical direction by 0.8, 1.8, 1.5, and 0.5 a.u., respectively. (e) Temperature dependence of the energy (top panel), integrated intensity (middle panel), and linewidth (bottom panel), extracted from the fits of the modes indicated by green triangles, gray circles, and orange squares in (c) and (d). The change in background color gradient from red to blue indicates the DW transition temperature.
    }
    \label{fig:2}
\end{figure*}

\subsection{Phononic Raman scattering} 

High-quality La$_4$Ni$_3$O$_{10}$ single crystals with the $P2_1/a$ structure (Figs.~\ref{fig:1}a,b) were grown using the optical floating-zone method. A transition in magnetic susceptibility measurements was observed at $T \approx 140$\,K. For further details about the sample synthesis and characterization, see Methods and Supplemental Material Sec. I and Sec. II~\cite{suppmat}.  

La$_4$Ni$_3$O$_{10}$ in the monoclinic $P2_1/a$ space group ($C_{2h}$ point group) exhibits 99 vibrational modes at the $\Gamma$ point, out of which 51 are infrared active (26 $A_u$ + 25 $B_u$) and 48 are Raman active (24 $A_g$ + 24 $B_g$). The corresponding Raman tensors are given in the Supplemental Material, Sec. III~\cite{suppmat}. The relation between the $x$, $y$, and $z$ directions of the laboratory frame of the experiment and the $a$, $b$, and $c$ crystallographic directions is displayed in Figs.~\ref{fig:1}a,b. Note that the monoclinic angle $\beta$ introduces an offset of $\sim$$11^\circ$ between the $c$- and $z$-directions. 

Figure~\ref{fig:1}c provides an initial overview on the temperature dependence of the Raman spectra of a La$_4$Ni$_3$O$_{10}$ single crystal. The scattering geometry including the photon polarizations are given in Porto's notation, 
where $u$ indicates that no analyzer was inserted in the scattered light path, \textit{i.e.}, $\bar z(x'u)z$ contains both $\bar z(x'x')z$ and $\bar z(x'y')z$ scattering contributions. In the following, we omit the labels $\bar z$ and $z$, denoting the incident and scattered photon directions in backscattering geometry, respectively, for simplicity. For technical details, see Methods. 

In the unpolarized ($x'u$) spectrum at 334\,K in Fig.~\ref{fig:1}c, distinct phonon modes appear in the ranges 60–180\,cm$^{-1}$ and 350–450\,cm$^{-1}$, alongside a broad hump centered around 910\,cm$^{-1}$ (shaded area in Fig.~\ref{fig:1}c). Upon cooling, the phonon linewidths sharpen significantly and additional modes emerge between 60 and 515\,cm$^{-1}$, while the broad hump shows minor temperature dependence. The most pronounced spectral changes occur between 152\,K and 120\,K, indicating that the DW transition at $T_\mathrm{DW} \approx 140$\,K affects the Raman scattering.

For a more detailed analysis, we focus in Fig.~\ref{fig:2} on polarized spectra and features below 600\,cm$^{-1}$, which is the typical energy range of phonons in other nickelates \cite{Hepting2014} and related oxides \cite{Devereaux2007,Liu1999}. Figures~\ref{fig:2}a,b display Raman spectra in the $x'x'$, $xx$, $x'y'$, and $xy$ polarization channels, acquired at 315 and 50~K, respectively. At both temperatures, differences between the $x'x'$ and $xx$ spectra, as well as between the $x'y'$ and $xy$ spectra, are subtle. This is consistent with the relatively small in-plane (orthorhombic) asymmetry between the $a$ and $b$ directions in La$_4$Ni$_3$O$_{10}$ (see Supplemental Material, Sec. I~\cite{suppmat}). By contrast, the difference between spectra in the parallel and cross polarization channels ($x'x'$ and $x'y'$ or $xx$ and $xy$) is striking, with several modes appearing exclusively in one configuration. However, due to the monoclinic angle $\beta$ of La$_4$Ni$_3$O$_{10}$, the diagonal and off-diagonal elements of the Raman tensor mix in our employed backscattering geometry, leading to the appearance of both $A_g$ and $B_g$ modes in each channel. 

The temperature dependence of several representative phonon modes between 355 and 515\,cm$^{-1}$ is displayed in Figs.~\ref{fig:2}c,d. Complementary analyses of lower-energy phonons are presented in the Supplemental Material, Fig.~S3~\cite{suppmat}. In Fig.~\ref{fig:2}c, a mode centered around 355\,cm$^{-1}$ increases in intensity upon cooling from 334 to 152 K, followed by a strong decrease in intensity for temperatures below $T_\mathrm{DW}$ (green triangles in Fig.~\ref{fig:2}c). A mode around 400\,cm$^{-1}$ (gray circle in Fig.~\ref{fig:2}c) continuously sharpens upon cooling from 334 to 55 K, while it retains significant intensity even below $T_\mathrm{DW}$. A mode around 515\,cm$^{-1}$ (orange squares in Fig.~\ref{fig:2}d) possesses notable intensity only below $T_\mathrm{DW}$ whereas it is absent at higher temperatures. The energy, integrated intensity, and linewidth of these three phonons, extracted from fits using Voigt profiles (see Methods), are summarized in Fig.~\ref{fig:2}e. 

In order to assign the phonon modes with these distinct behaviors, we perform density functional theory (DFT)-based structural relaxations of the $P2_1/a$ structure of La$_4$Ni$_3$O$_{10}$, followed by phonon frequency calculations using density-functional perturbation theory (DFPT) in combination with the PHONOPY package (see Supplemental Material, Sec. IV~\cite{suppmat}). The computed phonon frequencies are displayed as black triangles in Fig.~\ref{fig:2}b, and a table of all frequencies is provided in the Supplemental Material, Table~S3~\cite{suppmat}. 
The computed range, spanning from 55 to 574\,cm$^{-1}$, agrees reasonably well with the experimentally observed phonons from 61 to 516\,cm$^{-1}$ at 50 K (Fig.~\ref{fig:2}b). Note that the $P2_1/a$ symmetry persists for La$_4$Ni$_3$O$_{10}$ below $T_\mathrm{DW}$ \cite{Kumar2020,Rout2020}, supporting a comparison with the 50 K spectrum.

The atomic displacement patterns of selected phonon modes are displayed in the Supplemental Material, Fig.~S4~\cite{suppmat}, including a low-energy mode around  130\,cm$^{-1}$ that involves Ni-O bending vibrations in the outer layers of the trilayer unit. Notably, this mode was identified as a coherent Raman-active phonon that can be excited by ultrashort laser pulses in Ref.~\cite{Li2025SDW}. The prominent phonon observed around 402\,cm$^{-1}$ in Fig.~\ref{fig:2}c corresponds likely to an $A_g$ mode involving  bond stretching vibrations of basal oxygen in the outer Ni-O planes. This vibration pattern is reminiscent of quadrupolar modes in La$_2$CuO$_4$, which, unlike the oxygen breathing modes, are insensitive to the on-site Coulomb repulsion $U_d$ of the Cu 3$d$ orbitals~\cite{Falter2002}. Alternatively, the observed phonon might correspond to $A_g$ modes computed at slightly lower energies that involve quadrupolar bond-stretching vibrations of the basal oxygen atoms in the inner Ni–O plane. The phonon that emerges below $T_\mathrm{DW}$ in Fig.~\ref{fig:2}d near 517\,cm$^{-1}$ likely corresponds to a computed $B_g$ mode that is characterized by a breathing distortion of the inner oxygen octahedra.

In regions of dense phonon clustering in Fig.~\ref{fig:2}b, individual modes are difficult to distinguish, especially between 200 and 325\,cm$^{-1}$ in the parallel polarization channels. While a high density of phonon modes in this range is consistent with our calculations (see Supplemental Material, Table~S3~\cite{suppmat}), the overlap of peaks hinders the unambiguous assignment of individual modes.

\begin{figure}
    \centering
    \includegraphics[width=1\linewidth]{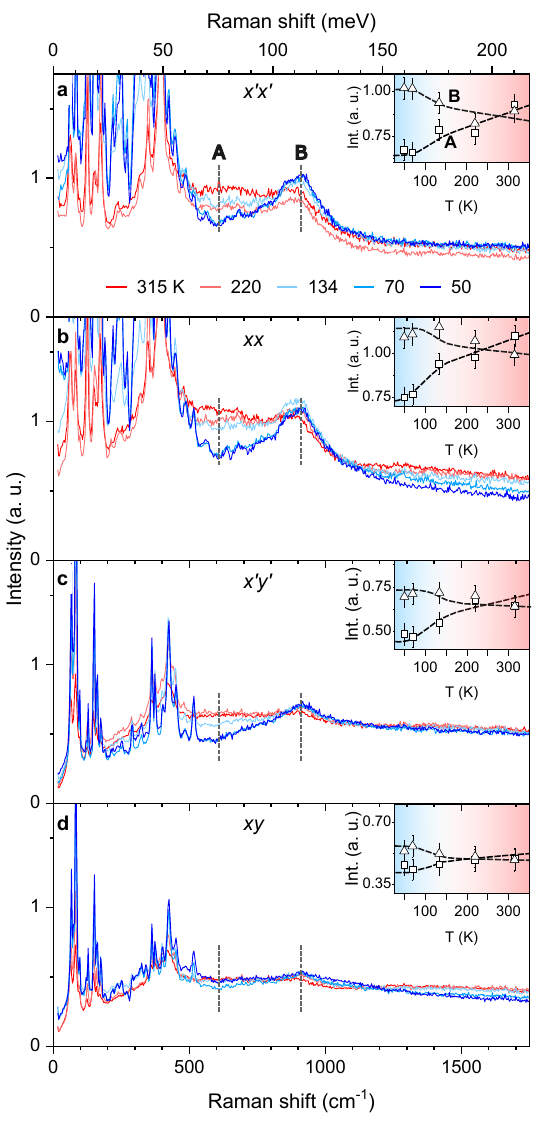}
    \caption{Electronic Raman scattering in La$_4$Ni$_3$O$_{10}$. (a-d) Raman spectra across a wide energy range, acquired at different temperatures (red to blue lines) in $x'x'$, $xx$, $x'y'$, and $xy$ scattering configuration (top to bottom panel), respectively. Vertical dashed lines at 610\,cm$^{-1}$ (label $A$) and 910\,cm$^{-1}$ (label $B$) indicate regions in the spectra with a distinct temperature dependence. The insets show the corresponding temperature dependence (square symbols for $A$, triangle symbols for $B$). Dashed lines in the insets serve as a guide to the eye. 
    }
    \label{fig:5}
\end{figure}

\begin{figure*}[tb]
    \centering
    \includegraphics[width=1\linewidth]
    {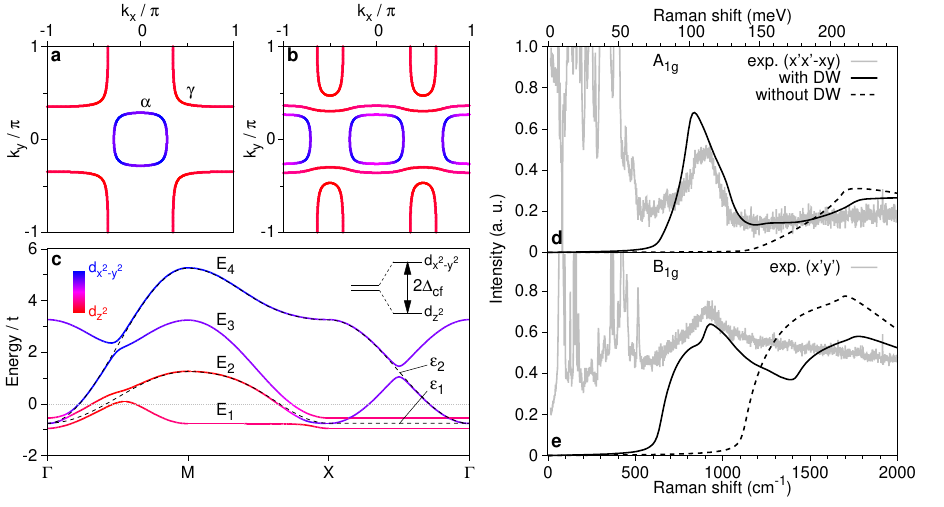}
    \caption{Raman response in the two-orbital model with density wave.
    (a) Fermi surface in the two-orbital model with $\alpha$ and $\gamma$ sheets. The color code represents $d_{x^2-y^2}$ (blue), $d_{z^2}$ (red), or mixed (magenta) orbital character.
    (b) Fermi surface in the presence of an uniaxial DW with ${\bf q} = (\pi,0)$ and $\delta = 0.2 t$ (see text).
    (c) Energy bands $E_1$ to $E_4$ of the two-orbital model in presence of the DW, with the orbital character resolved according to the color code. For comparison, the bands $\varepsilon_1$  (lower black dashed line) and $\varepsilon_2$ (upper black dashed line) in the absence of a DW  are also shown. The inset illustrates the splitting $2\Delta_\mathrm{cf}$ between the $d_{x^2-y^2}$ and $d_{z^2}$ orbital levels.
    (d) Raman response in the $A_{1g}$ channel, calculated in the presence (black solid line) and absence of the DW (black dashed line). An energy-dependent quasiparticle weight was applied
  (see Methods section). Experimental data (gray line) from the $x'x' - xy$ channel are superimposed. 
    (e) Raman response in the $B_{1g}$ channel, calculated in the presence (black solid line) and absence of the DW (black dashed line). Experimental data (gray line) from the $x'y'$ channel are superimposed. 
    }
    \label{fig:FS}
\end{figure*}

\subsection{Electronic Raman scattering} 

After analyzing the phonon sector across the DW transition, we now turn to an extended energy range. In the spectra up to 1750\,cm$^{-1}$ (Fig.~\ref{fig:5}a), a broad peak centered around 910\,cm$^{-1}$ emerges most prominently in the $x'x'$ and $xx$ channels. Its linewidth exceeds 150\,cm$^{-1}$ even at the lowest measured temperatures, which is substantially broader than that of the phonon modes observed up to 515\,cm$^{-1}$.  Notably, this feature appears to divide the Raman spectrum into two regimes: a high-energy region that is nearly flat above $\sim$1140\,cm$^{-1}$ and exhibits weak temperature dependence, and a lower-energy region with strong temperature dependence and regions of either increasing or depleting spectral weight upon cooling. This separation of the Raman spectra into distinct energy scales by the 910\,cm$^{-1}$ peak suggests that it is likely associated with an electronic excitation rather than multi-phonon scattering. Furthermore, the 910\,cm$^{-1}$ peak and a nearly flat background above $\sim$1140\,cm$^{-1}$ appear similarly in spectra acquired with both 632.8 nm (Fig.~\ref{fig:5}) and 532 nm (see Supplemental Material, Sec.~V~\cite{suppmat}) laser excitation, confirming the robustness of the features. 

The polarization dependence provides additional information on the symmetry of the electronic Raman response. The insets in Figs.~\ref{fig:5}a-d display the temperature dependence of the Raman intensity at two representative energies (610 and 910\,cm$^{-1}$) in the four polarization channels. In the $x'x'$ channel, the intensity at 610\,cm$^{-1}$ (Fig.~\ref{fig:5}a, label $A$) decreases upon cooling, particularly below $T_\mathrm{DW}$. This trend reflects a broader suppression of spectral weight in the surrounding energy range at least up to 700\,cm$^{-1}$. Below 500\,cm$^{-1}$ the detailed behavior of any broad spectral weight distribution is masked by the rapidly growing phonon intensity. In contrast to the spectral weight depletion, the 910\,cm$^{-1}$ peak gains intensity with cooling (Fig.~\ref{fig:5}a, label $B$), most strongly near $T_\mathrm{DW}$ in the $x'x'$ channel. The contrasting temperature evolution of features $A$ and $B$ is similarly observed in the other polarization channels (insets in Fig.~\ref{fig:5}), although it is almost absent in the $xy$ channel.

\subsection{Density wave in a two-orbital model} 

To gain insights into the origin of the 910\,cm$^{-1}$ feature and the temperature-dependent spectral weight at energies lower than 910\,cm$^{-1}$ (Fig.~\ref{fig:5}), we next develop a minimal model to compute the electronic Raman response of La$_4$Ni$_3$O$_{10}$.
In constructing our model, it is crucial to take into account the multi-band and multi-orbital nature of La$_4$Ni$_3$O$_{10}$, in contrast to effective single-band descriptions, as often used for cuprates~\cite{Lee_review2006,Devereaux2007}.
Specifically, first-principles calculations indicate that the active electronic states in La$_4$Ni$_3$O$_{10}$ near the Fermi level are composed of $d_{x^2-y^2}$ and $d_{z^2}$ orbitals \cite{Chen2024Trilayer,Yang2024}, suggesting that at least two orbitals are
required to capture the essential low-energy physics and associated electronic excitations. 
Since both the charge- and spin-density modulations in La$_4$Ni$_3$O$_{10}$ occur primarily within the NiO$_2$ planes \cite{zhang2020intertwined}, we restrict our analysis to a two-dimensional model with momentum $\mathbf{k}=(k_x,k_y)$ and in-plane hopping, while retaining the two-orbital local basis. In addition, we adopt a simplified fourfold symmetric in-plane geometry for La$_4$Ni$_3$O$_{10}$, which allows for a decomposition of the Raman response into the conventional $A_{1g}$, $B_{1g}$, and $B_{2g}$ symmetry channels, in analogy to square-lattice cuprates with CuO$_2$ planes. This approximation is compatible with the observed similarity of the phonon spectra in the $x'x'$ and $xx$ as well as the $x'y'$ and $xy$ channels in Figs.~\ref{fig:2}a,b, which would not be expected in systems with pronounced orthorhombicity. 

We define the model Hamiltonian as $H = H_{0} + H_\mathrm{DW}$, where $H_{0}$ describes electrons hopping within and between $d_{x^2-y^2}$ and $d_{z^2}$ orbitals (for details of the model see Methods). 
The corresponding Raman vertices are obtained in the orbital basis and are therefore matrices in the $d_{x^2-y^2}$/$d_{z^2}$ orbital space, rather than scalar band-curvature factors.
The crystal field $2\Delta_\mathrm{cf}$ splits the $e_g$ orbital manifold into two levels (inset in Fig.~\ref{fig:FS}c), yielding dispersive bands $\varepsilon_1$ and $\varepsilon_2$  (black dashed lines in Fig.~\ref{fig:FS}c) and two FS sheets, comprising a circular electronlike pocket with $d_{x^2 - y^2}$ and minor $d_{z^2}$ character around $\Gamma$, and a holelike pocket of predominantly $d_{z^2}$ character centered at $(\pi, \pi)$ (Fig.~\ref{fig:FS}a).
This FS qualitatively reproduces the topology and orbital character reported by ARPES and first-principles calculations for La$_4$Ni$_3$O$_{10}$~\cite{Yang2024,Li2017,Chen2024Trilayer}, including the $\alpha$ and $\gamma$ sheets as denoted in Refs.~\cite{Yang2024,Chen2024Trilayer}. The $\beta$ sheet, with trilayer-induced splitting into $\beta$ and $\beta'$ and mixed $d_{x^2 - y^2}$/$d_{z^2}$ character~\cite{Yang2024,Chen2024Trilayer}, is not implemented in our minimal model.

The term $H_\mathrm{DW}$ effectively describes a density modulation within the $d_{x^2-y^2}$ and $d_{z^2}$ orbitals, characterized by a wave vector ${\bf q}$ and the parameter $\delta$, which modulates the chemical potential to simulate the presence of a DW (see Methods). We consider a uniaxial density modulation with ${\bf q} = (\pi,0)$ and $\delta = 0.2 t$, using an effective hopping parameter $t = 0.28$~eV, reflecting effective mass renormalization due to correlations in La$_4$Ni$_3$O$_{10}$. The value of $\delta$ is chosen to reproduce the experimentally relevant DW gap scale, rather than being predicted from first principles. The resulting FS is plotted in Fig.~\ref{fig:FS}b and the energy bands $E_1$ to $E_4$, with their orbital character resolved, are displayed in Fig.~\ref{fig:FS}c.
Furthermore, we introduce an energy-dependent quasiparticle weight (see Methods)
which effectively reduces the intensities of high-energy interband transitions.

Figure~\ref{fig:FS}d shows the computed Raman response in the $A_{1g}$ channel in the presence ($\delta = 0.2 t$) and absence ($\delta=0$) of a DW. A pronounced peak emerges around 910\,cm$^{-1}$ when the DW is present. The $B_{1g}$ response exhibits qualitatively similar behavior, with a DW-induced peak of comparable intensity, albeit with a distinct lineshape (Fig.~\ref{fig:FS}e). 
Considering a simplified fourfold symmetric in-plane geometry of La$_4$Ni$_3$O$_{10}$, with the principal axes aligned along the Ni--O--Ni bond directions rather than the monoclinic setting in Fig.~\ref{fig:1}b, the $A_{1g}$ and $B_{1g}$ channels correspond to the $x'x' - xy$ and $x'y'$ polarization configurations in Fig.~\ref{fig:5}, respectively. The corresponding Raman data, overlaid in Figs.~\ref{fig:FS}d,e, show good agreement with the calculated responses. 

The computed Raman response in the absence of a DW term does not exhibit a peak near 910\,cm$^{-1}$ in either the $A_{1g}$ or $B_{1g}$ channel in Figs.~\ref{fig:FS}d,e. Instead, only a broad onset with spectral weight from interband transitions appears around 1140\,cm$^{-1}$. Given the values of $\Delta_\mathrm{cf}$, $t$, and $\delta$ used in our model calculation, a DW gap of $\Delta_\mathrm{DW} = 56$\,meV follows, thus strongly suggesting that the experimentally observed feature around 910\,cm$^{-1}$ ($\approx 114$\,meV) 
originates from the DW and that the gap magnitude of La$_4$Ni$_3$O$_{10}$ is $2\Delta_\mathrm{DW} \approx 114$\,meV.

In our minimal model, we neglect second-nearest-neighbor hopping $t'$, which is much smaller than nearest-neighbor hopping $t$ in La$_4$Ni$_3$O$_{10}$. The $B_{2g}$ Raman vertex, which scales with $t'^2$, is therefore vanishingly small, and a computed $2\Delta_{\mathrm{DW}}$ peak is expected to be minor in the $B_{2g}$ channel, consistent with the faint temperature-dependent effects observed for electronic Raman scattering in $xy$ polarization in Fig.~\ref{fig:5}d.

To rule out a single-orbital origin of the $2\Delta_{\mathrm{DW}}$ peak, we next compute the Raman response restricting the Raman vertex to either the $d_{x^2-y^2}$ or the $d_{z^2}$ orbital sector (see Supplemental Material, Sec.~VI~\cite{suppmat}). Notably, these single-orbital projections do not reproduce the experimentally observed intensity ratio getween the $2\Delta_{\mathrm{DW}}$ peak in the $A_{1g}$ and $B_{1g}$ channels as well as the linewidths and lineshapes. Agreement with experiment is obtained only when both orbital contributions and their mixing are retained, demonstrating that the Raman-active DW excitation has multiorbital character.

\section{Discussion}

Our Raman results in the phonon sector (Fig.~\ref{fig:2}) reveal that the lattice dynamics of La$_4$Ni$_3$O$_{10}$ are affected by the DW transition at $T_\mathrm{DW} \approx 140$\,K. Phonon anomalies near $T_\mathrm{DW}$ were also reported in previous Raman studies~\cite{Li2025SDW,Gim2025,SoniaRaman}, although the observed phonon intensity changes in our experiment appear comparatively more pronounced (Fig.~\ref{fig:1}c and Figs.~\ref{fig:2}a-e). Moreover, the emergent clusters of densely spaced phonons between 200 and 325\,cm$^{-1}$ (see $x'x'$ and $xx$ in Fig.~\ref{fig:2}b) are not apparent in Ref.~\cite{Gim2025}. On the other hand, Refs.~\cite{Gim2025,SoniaRaman} reported strong phonon peaks at $\sim$570 and $\sim$688\,cm$^{-1}$, respectively, which are absent in our spectra. In particular, the $\sim$688\,cm$^{-1}$ peak lies far beyond the cut-off frequency of our phonon calculation for La$_4$Ni$_3$O$_{10}$ in the monoclinic $P2_1/a$ structure, possibly indicating a deviating structure due to presence of oxygen vacancies and/or intergrown secondary phases, such as other RP or polymorph phases~\cite{Chen2024JACS,Puphal2024}, in the samples of Refs.~\cite{Gim2025,SoniaRaman}. Orthorhombic~\cite{zhang2020,Yuan2024} and non-superconducting tetragonal~\cite{Shi2025} variants of La$_4$Ni$_3$O$_{10}$ might also yield Raman spectra with features deviating from Fig.~\ref{fig:1}c and Figs.~\ref{fig:2}a,b.

Among the most striking fingerprints of the DW transition in the phonon sector is the emergence of new features below $T_\mathrm{DW}$, including the mode around 515\,cm$^{-1}$ (orange squares in Figs.~\ref{fig:2}d,e) and the clusters  between 200 and 325\,cm$^{-1}$ (Fig.~\ref{fig:2}b). Nevertheless, we refrain from a definitive assignment of their origin to zone-folding induced by the DW, as our DFT-based phonon calculation for the ordinary $P2_1/a$ unit cell already predicts modes at these energies (triangle symbols in Fig.~\ref{fig:2}b). Instead, their appearance may stem from reduced metallic screening below the metal-to-metal transition at $T_\mathrm{DW}$~\cite{zhang2020intertwined,Li2017} due to partial FS gapping, rendering previously overdamped $P2_1/a$ phonons observable.

Renormalization effects on the phonon self-energy, affecting both the real (frequency) and imaginary (linewidth) parts, are relatively subtle in our data compared to those of zone-center phonons of canonical DW systems~\cite{Pouget2024}. The linewidth narrowing with decreasing temperature of the phonons at 355 and 400\,cm$^{-1}$ (Fig.~\ref{fig:2}e, bottom panel) is consistent with anharmonic phonon decay, without influence from a phase transition, and slight anomalies near $T_\mathrm{DW}$ in our data remain within error bars. For the phonon frequencies (Fig.~\ref{fig:2}e, top panel), we observe a subtle yet statistically significant dip below $T_\mathrm{DW}$, interrupting an otherwise monotonic trend of anharmonic hardening. A similar dip occurs for the low-energy bending modes at 129 and 176\,cm$^{-1}$ (see Supplemental Material, Fig.~S3~\cite{suppmat}), involving La, Ni, and O displacements. While we cannot rule out that the frequency dip originates from a change of electron-phonon coupling across $T_\mathrm{DW}$, the faint softening of various low- and high-energy modes, together with the diminishing of intensity of a few specific modes (orange squares in Figs.~\ref{fig:2}d,e), more likely reflects subtle structural changes, such as $b$-axis expansion\cite{Kumar2020} and small tilt-angle variations, while $P2_1/a$ symmetry is retained~\cite{Kumar2020,Rout2020}.

We also note that all resolved modes exhibit symmetric peak profiles both above and below $T_\mathrm{DW}$ (Figs.~\ref{fig:2}a-d and Supplemental Material, Sec.~V~\cite{suppmat}), without Fano lineshapes that would signal coupling between a discrete phonon excitation and the electronic continuum~\cite{Bakr2009}. Both the absence of clear self-energy renormalization and Fano lineshapes agrees with theoretical predictions of weak to moderate electron–phonon coupling in RP nickelates~\cite{Zhan2025,Zhu2025,Ouyang2024,Kumar2020}. Furthermore, this finding aligns with recent inelastic x-ray scattering (IXS) results on Pr$_4$Ni$_3$O$_{10}$, where no phonon anomalies were detected at the wavevectors associated with either CDW or SDW order~\cite{Jia2025}, suggesting only a minor role of electron–phonon interactions in the DW transition. This contrasts with an unpolarized Raman study on La$_4$Ni$_3$O$_{10}$ powder~\cite{SoniaRaman}, which proposed a crossover from an electron--phonon coupled high-temperature regime to an electron--electron correlated phase below $T_\mathrm{DW}$.

A second energy scale, extending beyond the phonon sector up to $\sim$114\,meV ($\sim$910\,cm$^{-1}$), is characterized by temperature-dependent continuum spectral weight, with its upper bound marked by the DW peak. The distinct electronic Raman responses observed in different polarization channels (Fig.~\ref{fig:5}) show that the DW-related spectral-weight redistribution is strongly symmetry dependent. In a multiorbital system, this polarization dependence calls for an analysis based on the orbital-matrix structure of the Raman vertex~\cite{Blesio2024}, rather than an interpretation solely in terms of selected Brillouin-zone regions. Combined with explicit two-orbital model calculations of the Raman response, this enables identification of the multiorbital character of the Raman-active DW excitation.

Importantly, our calculated $2\Delta_\mathrm{DW}$ peaks in the $A_{1g}$ and $B_{1g}$ channels (Figs.~\ref{fig:FS}d,e) align closely with experiment, without requiring channel-dependent scaling factors, underscoring the consistency of our model. The irregular peak lineshapes, including the $\sim$95\,meV ($\sim$760\,cm$^{-1}$) shoulder in $B_{1g}$, are also reproduced. Complementary calculations, where the Raman vertex accounts exclusively for either $d_{x^2 - y^2}$ to $d_{x^2 - y^2}$ or $d_{z^2}$ to $d_{z^2}$ transitions, yield irregular lineshapes as well (see Supplemental Material, Sec.~VI~\cite{suppmat}), implying that lineshape features of the $2\Delta_\mathrm{DW}$ peak are primarily governed by the FS topology rather than distinct orbital contributions. However, without orbital mixing, the calculated relative intensity between the $2\Delta_\mathrm{DW}$ peaks in the $A_{1g}$ and $B_{1g}$ channels as well as the linewidths deviate from experiment (see Supplemental Material, Sec.~VI~\cite{suppmat}), corroborating the multiorbital nature of the Raman-active DW excitation. These findings challenge the orbital-selective $d_{x^2 - y^2}$ assignment of Ref.~\cite{Gim2025}, whereas they align with theoretical studies emphasizing that a multiorbital basis is required for modeling the intertwined electronic phases and instabilities in RP nickelates~\cite{Chen2024Trilayer,Yang2024,Zhang2024b,LaBollita2024b,Zhang2024a,Zhang2025b,Lu2025,Zhang2025d}.

As a key result of our study, the position of the $2\Delta_\mathrm{DW}$ peak around $\sim$114\,meV marks the energy gap in La$_4$Ni$_3$O$_{10}$. This value is comparable to that determined by ultrafast optical spectroscopy on La$_4$Ni$_3$O$_{10}$~\cite{Xu2025,Xu2025Pressure}, but deviates substantially from the lower range of reported values, with $\Delta_\mathrm{DW}$ between 12 and 35\,meV from ARPES, STM, and a different Raman study~\cite{Li2017,du2024,Li2025STM,SoniaRaman}. Notably, optical conductivity measurements on bilayer La$_3$Ni$_2$O$_7$ extracted $2\Delta_\mathrm{CDW} \approx 100$\,meV and assigned this gap unambiguously to the CDW, since the two ordering phenomena are disjoint in that material, with $T_\mathrm{CDW} \approx 110$\,K and $T_\mathrm{SDW} \approx 150$\,K~\cite{Liu_2024}. The similarity in gap magnitudes suggests that the $\sim$114\,meV gap in La$_4$Ni$_3$O$_{10}$ is also associated with the CDW, while the lower-energy features~\cite{Li2017,du2024,SoniaRaman} may correspond to the SDW gap. In our spectra, the low-energy region is partly obscured by clustered phonons (Fig.~\ref{fig:5}), although we note enhanced spectral weight below $\sim$50 meV in the $xx$ channel below $T_\mathrm{DW}$, warranting future high-resolution Raman work to clarify a possible hierarchy of DW-related energy scales.

The temperature dependence of the $2\Delta_\mathrm{DW}$ peak at 114\,meV and the continuum spectral weight at lower energies exhibit several unconventional characteristics: Although the peak gains most spectral weight below $T_\mathrm{DW}$ (insets in Fig.~\ref{fig:5}, label $B$), it is already observable above the transition temperature. This  deviates from expectations based on a BCS mean-field scenario, where both the gap and the associated collective mode should vanish at $T = T_\mathrm{DW}$. Instead, the persistence suggests that the CDW amplitude is already finite above $T_\mathrm{DW}$, but its correlation length short-ranged and quasi-two-dimensional. In addition, the continuum depletes gradually with cooling (insets in Fig.~\ref{fig:5}, label $A$) rather than a ``coherent" opening of the gap, reminiscent of pseudogap phenomena in cuprates and other correlated materials~\cite{Devereaux2007,Keimer2015,Uchida2011}. An incoherent gap opening has also been found in optical conductivity measurements on La$_3$Ni$_2$O$_7$~\cite{Liu_2024}, as well as in related oxides~\cite{Liu1999}.
Notably, recent Landau theory considerations predict that the CDW is subordinate to the SDW~\cite{Norman2025}, and the small 12\,meV gap in La$_4$Ni$_3$O$_{10}$ observed in ARPES appears to follow a conventional BCS-like temperature dependence~\cite{du2024}.

While our minimal model is not devised to capture the full complexity of La$_4$Ni$_3$O$_{10}$, such as the three-dimensionality of the DW order~\cite{Li2017}, the presence of four FS sheets~\cite{Chen2024Trilayer}, and structural subtleties~\cite{zhang2020,Yuan2024,Shi2025}, it nonetheless yields several robust and general insights: (i) The identification of the multiorbital character of the DW does not depend on details of the underlying FS. Specifically, our model considers one electronlike and one holelike pocket, representing the $\alpha$ and $\gamma$ FS sheets of La$_4$Ni$_3$O$_{10}$, with hybridized $d_{x^2 - y^2}$/$d_{z^2}$ and nearly pure $d_{z^2}$ orbital character, respectively. Nevertheless, similar multiorbital DW features in the Raman response can be obtained from a model that mimics the $\alpha$ and a $\beta$-like sheet of La$_4$Ni$_3$O$_{10}$, given that the $\beta$ sheet includes at least partial $d_{z^2}$ character (not shown here). (ii) The results are robust against the choice of the exact wave vector of the DW, as long as it connects FS regions with mixed orbital character. (iii) An upper bound for the CDW amplitude can be estimated: we obtain a modulation of $\pm 0.2$ electrons relative to the mean, given that the canonical electron filling $n = 4/3$ and a modulation $\delta = 0.2 t$ of the chemical potential yields good agreement with the experimental data in Fig.~\ref{fig:FS}d,e, and assuming that $\Delta_\mathrm{CDW}$ cannot exceed the interband transitions at $\Gamma$ in a fully coherent‐band picture.

\section{Conclusion}

In summary, our polarization-resolved Raman study uncovers the phononic hallmarks of the DW transition in La$_4$Ni$_3$O$_{10}$, including the emergence of modes below $T_\mathrm{DW}$, intensity changes of persisting modes, and subtle frequency anomalies. The combination of momentum selectivity of electronic Raman scattering and two-orbital model calculations enables the identification of the $2\Delta_\mathrm{DW}$ peak of La$_4$Ni$_3$O$_{10}$ and its multiorbital origin, involving both Ni-$d_{x^{2}-y^{2}}$ and $d_{z^{2}}$ states. Spectral weight within the DW gap of approximately 114\,meV depletes incoherently with decreasing temperature, while the  $2\Delta_\mathrm{DW}$ peak persists even above $T_\mathrm{DW}$. The corresponding ratio $2\Delta_\mathrm{DW}/k_\mathrm{B}T_\mathrm{DW}$ exceeds the weak-coupling BCS limit, signaling short-range correlations and a strong-coupling character of the DW instability.  

Notably, our results lend strong support to theoretical models of RP nickelates that incorporate multiorbital DW instabilities, particularly those that assign prominent roles to $d_{z^{2}}$ states~\cite{Yang2024,Zhang2024b,LaBollita2024b,Zhang2024a,Zhang2025b,Lu2025,Zhang2025d}. Given that both the orbital character and momentum structure of the hybridized Ni-$d_{x^{2}-y^{2}}$/$d_{z^{2}}$ DW correlations identified here in the ordered phase at ambient pressure exhibit close correspondence to the persisting DW fluctuations at high pressure, several constraints on superconducting pairing models can be inferred. Specifically, if these fluctuations either compete with or mediate superconductivity, their finite wavevector and interorbital nature might favor pairing states with sign-changing order parameters between distinct FS sheets, supporting extended $s^\pm$-wave pairing symmetry as a viable candidate~\cite{Zhang2024b,Yang2024,Zhang2024a,Huang2024,Fan2025}, and strongly motivating future Raman investigations not only of the pressure dependence of the DW phenomenon, but also the potential appearance of a superconducting pair-breaking peak in RP nickelates.


\section{Methods}

\subsection{Samples} 

High-quality La$_4$Ni$_3$O$_{10}$ single crystals were grown using the optical floating-zone method and post-annealed in oxygen atmosphere to optimize oxygen stoichiometry. Single-crystal x-ray diffraction (SXRD) revealed the monoclinic $P2_1/a$ structure (space group no.~14). A magnetic transition was observed in susceptibility measurements at $T \approx 140$\,K. Details about the sample synthesis and characterization are given in the Supplemental Material, Sec.~I~\cite{suppmat}. Phase purity and trilayer stacking in the La$_4$Ni$_3$O$_{10}$ crystals was confirmed by scanning transmission electron microscopy (see Supplemental Material Sec.~II~\cite{suppmat}). The transition temperature $T_\mathrm{DW} \approx 140$~K is consistent with several previous reports of the concomitant SDW and CDW ordering in monoclinic La$_4$Ni$_3$O$_{10}$~\cite{zhang2020intertwined,Kakoi2024,Li2025STM,Khasanov2025,Fukamachi2001}, where neutron diffraction established that the SDW emerges in the outer NiO$_6$ layers of the trilayer block, while SXRD indicated that the CDW modulation occurs in each of the three layers~\cite{zhang2020intertwined}. 

In transport measurements on crystals from the same batch used for this Raman study, a resistivity downturn was observed for pressures above $\sim$20\,GPa (to be published elsewhere), suggesting that our crystals exhibit high-pressure superconductivity consistent with La$_4$Ni$_3$O$_{10}$ samples in Refs.~\cite{Zhu2024,Sakakibara2024,Li_2024HTS4310,Zhang2025}. By contrast, for an orthorhombic variant, which likely corresponds to heavily oxygen-deficient La$_4$Ni$_3$O$_{10-\delta}$, the DW transition temperature deviates from 140\,K~\cite{zhang2020,Yuan2024}, and a tetragonal variant showed neither a DW transition at ambient pressure nor superconductivity under high pressure~\cite{Shi2025}.

\subsection{Polarized Raman scattering} 

The Raman experiments were performed with a HORIBA Jobin Yvon LabRAM HR800 spectrometer in backscattering geometry, using a He-Ne laser (wavelength 632.8 nm) focused on the sample with a x50 magnification ultra-long working distance objective. The laser power under the microscope was less than 1~mW to minimize heating effects on the sample. The temperatures indicated in the figures of the main text were determined by Stokes/anti-Stokes intensity analysis. All plotted Raman spectra were divided by the Bose-Einstein factor. An ultra-low frequency filter (consisting of two volume Bragg-grating filters) was used to suppress elastically scattered light. For the low-temperature measurements, a KONTI helium-flow cryostat from CryoVac was used.

The Raman spectra in Figs.~\ref{fig:2}c,d were fitted to a superposition of Voigt profiles resulting from a convolution of the intrinsic Lorentzian lineshape with the Gaussian spectrometer resolution, determined as 2\,cm$^{-1}$ from the spectrum of a Ne gas discharge lamp. For the background in the fits of Figs.~\ref{fig:2}c,d, hyperbola and linear functions were used, respectively.

\subsection{Phonon calculations} 

Details on the phonon calculations and a table of the computed Raman phonon frequencies are provided in the Supplemental Material, Sec. IV~\cite{suppmat}.

\subsection{Two-orbital model}

The minimal model with two orbitals is defined as
\begin{equation}
H = H_0 + H_{\mathrm{DW}},
\end{equation}
where the bare model without DW modulation is given by
\begin{equation}
H_0=\sum_{\mathbf{k}} \psi_{\mathbf{k}}^\dagger H_{\mathbf{k}} \psi_{\mathbf{k}},
\qquad
\psi_{\mathbf{k}}^\dagger=
\left(
x_{\mathbf{k}}^\dagger,\,
z_{\mathbf{k}}^\dagger
\right),
\end{equation}
where $x_{\mathbf{k}}^\dagger$ and $z_{\mathbf{k}}^\dagger$ create electrons in the $d_{x^2-y^2}$ and $d_{z^2}$ orbitals, respectively. The $2\times 2$ matrix $H_{\mathbf{k}}$ is the Hamiltonian in the orbital basis, with
\begin{equation}
H_{\mathbf{k}}=
\begin{bmatrix}
t_{xx}(\mathbf{k}) & t_{xz}(\mathbf{k}) \\
t_{xz}(\mathbf{k}) & t_{zz}(\mathbf{k})
\end{bmatrix},
\label{eq:Hk}
\end{equation}
where the element
\begin{align}
t_{xx}(\mathbf{k}) &= -\frac{3}{2} t \left(\cos k_x+\cos k_y\right)+\Delta_{\mathrm{cf}}-\mu,
\end{align}
describes hopping between $d_{x^2-y^2}$ orbitals,
\begin{align}
t_{zz}(\mathbf{k}) &= -\frac{1}{2} t \left(\cos k_x+\cos k_y\right)-\Delta_{\mathrm{cf}}-\mu,
\end{align}
corresponds to hopping between $d_{z^2}$ orbitals, and
\begin{align}
t_{xz}(\mathbf{k}) &= \frac{\sqrt{3}}{2} t \left(\cos k_x-\cos k_y\right),
\end{align}
accounts for hopping between the $d_{x^2-y^2}$ and $d_{z^2}$ orbitals. The crystal-field splitting of the $e_g$ orbital levels is denoted by $\Delta_{\mathrm{cf}}$, $\mu$ is the chemical potential, and $\mathbf{k}$ is the momentum in a two-dimensional square lattice. Thus, the model is two-dimensional in real and momentum space, while it is multiorbital in its local electronic basis.

The Hamiltonian in the band basis is diagonal
\begin{equation}
U_{\mathbf{k}}^\dagger H_{\mathbf{k}} U_{\mathbf{k}}
=
H_{\mathbf{k}}^{B}
=
\begin{bmatrix}
\varepsilon_1(\mathbf{k}) & 0 \\
0 & \varepsilon_2(\mathbf{k})
\end{bmatrix},
\label{eq:HkB}
\end{equation}
with band energies
\begin{equation}
\varepsilon_{1,2}(\mathbf{k})
=
\frac{t_{xx}(\mathbf{k})+t_{zz}(\mathbf{k})}{2}
\pm
\frac{r(\mathbf{k})}{2},
\end{equation}
where
\begin{equation}
r(\mathbf{k})
=
\sqrt{[t_{xx}(\mathbf{k})-t_{zz}(\mathbf{k})]^2
+
4t_{xz}^2(\mathbf{k})
}.
\end{equation}
In the numerical calculations, we use the effective hopping parameter $t = 0.28$~eV,  which is somewhat smaller than predicted by first principle calculations, but can be justified when considering the correlated character of La$_4$Ni$_3$O$_{10}$, increasing the effective mass. We fix $\Delta_\mathrm{cf} = t$, $\mu = -1.25 t$, and the density $n = 4/3$. 

Within the nonresonant effective-mass approximation, the Raman vertices are obtained from second derivatives of the low-energy Hamiltonian with respect to momentum. Since the model is multiorbital, the Raman vertex is a matrix in orbital space,
\begin{align}
R_{\bf k}^\gamma &= \left(\frac{\partial^2}{\partial k_x^2} \pm \frac{\partial^2}{\partial k_y^2}\right)H_{\bf k} 
\end{align}
\noindent where $+$ ($-$) corresponds to the $\gamma=A_{1g}$ ($B_{1g}$) vertex. Our model does not include second-nearest neighbor hopping $t'$, and therefore the $B_{2g}$ vertex is not considered. We therefore restrict the analysis to the $A_{1g}$ and $B_{1g}$ channels.

To describe the reconstructed phase, we introduce a DW potential with ordering vector $\mathbf{q}$,
\begin{equation}
H_{\mathrm{DW}}
=
\delta \sum_{\mathbf{k}}
\left(
x_{\mathbf{k}+\mathbf{q}}^\dagger x_{\mathbf{k}}
+
z_{\mathbf{k}+\mathbf{q}}^\dagger z_{\mathbf{k}}
+
\mathrm{H.c.}
\right),
\label{eq:HDW}
\end{equation}
where \(\delta\) is the amplitude of the DW potential, corresponding to a modulation of the local chemical potential $\mu$. In the calculations below we use $\delta=0.2\,t$ and $\mathbf{q}=(\pi,0)$, which is close to the dominant nesting vector of the model Fermi surface.

In the folded basis
\begin{equation}
\Phi_{\mathbf{k}}^\dagger=
\left(
x_{\mathbf{k}}^\dagger,\,
z_{\mathbf{k}}^\dagger,\,
x_{\mathbf{k}+\mathbf{q}}^\dagger,\,
z_{\mathbf{k}+\mathbf{q}}^\dagger
\right),
\end{equation}
the Hamiltonian becomes
\begin{equation}
H=\sum_{\mathbf{k}}
\Phi_{\mathbf{k}}^\dagger
\mathcal{H}_{\mathbf{k}}
\Phi_{\mathbf{k}},
\end{equation}
with
\begin{equation}
\mathcal{H}_{\mathbf{k}}
=
\begin{bmatrix}
H_{\mathbf{k}} & \delta I \\
\delta I & H_{\mathbf{k}+\mathbf{q}}
\end{bmatrix},
\label{eq:HkDW1}
\end{equation}
where $I$ is the $2\times 2$ identity matrix and the momentum sum is
over the entire Brillouin zone.
The reconstructed Hamiltonian is diagonalized by a $4\times 4$ unitary matrix $\mathcal{U}_{\mathbf{k}}$ to give four bands $\alpha$ with energies $E_{\alpha\mathbf{k}}$.

In the same folded orbital basis, the Raman vertex is approximated as block diagonal,
\begin{equation}
\mathcal{R}_{\mathbf{k}}^\gamma
=
\begin{bmatrix}
R_{\mathbf{k}}^\gamma & 0 \\
0 & R_{\mathbf{k}+\mathbf{q}}^\gamma
\end{bmatrix}.
\label{eq:RkDW}
\end{equation}
Transformation to the reconstructed band basis then yields
\begin{equation}
\mathcal{R}_{B,\mathbf{k}}^\gamma
=
\mathcal{U}_{\mathbf{k}}^\dagger
\mathcal{R}_{\mathbf{k}}^\gamma
\mathcal{U}_{\mathbf{k}}.
\end{equation}

The Raman response is evaluated as
\begin{equation}
\chi^\gamma(i\omega_n)
=
\sum_{\mathbf{k},\,i\nu_n}
\mathrm{Tr}
\left[
\mathcal{R}_{B,\mathbf{k}}^\gamma
\mathcal{G}_{\mathbf{k}}(i\nu_n+i\omega_n)
\mathcal{R}_{B,\mathbf{k}}^\gamma
\mathcal{G}_{\mathbf{k}}(i\nu_n)
\right],
\label{eq:chiMatsubara}
\end{equation}
where $\nu_n$ and $\omega_n$ are fermionic and bosonic Matsubara frequencies, respectively, and the trace is taken over the reconstructed band indices. Here $\mathcal{G}_{\mathbf{k}}$ is the quasiparticle part of the Green's function, \textit{i.e.}, we account for the Raman transitions between the coherent quasiparticle $\alpha$ bands. Due to correlations, an energy-dependent quasiparticle weight
$Z(E) \propto \eta^2 / (E^2 + \eta^2)$
with $\eta = 0.5 t$ is introduced, 
leading to \begin{equation}
\mathcal{G}_{\mathbf{k},\alpha\alpha'}(i\nu_n)
=
\delta_{\alpha\alpha'}\,
\frac{Z(E_{\alpha\mathbf{k}})}{i\nu_n-E_{\alpha\mathbf{k}}},
\end{equation}
where $E_{\alpha\mathbf{k}}$ are the eigenvalues of Eq.~\ref{eq:HkDW1}.

\section*{Data availability}
There are no publicly available research data or software supporting this manuscript. Requests for further information 
or data should be sent to the authors.

 \begin{acknowledgements}
We thank Y.~Gallais, G.~Blesio, C.~Falter, M.~J.~Graf von Westarp, and A.~von Ungern-Sternberg Schwark for insightful discussions. X.W. is supported by the National Key R\&D Program of China (Grant No. 2023YFA1407300) and the National Natural Science Foundation of China (Grants No. 12574151, No. 12447103, and No. 12447101). M.B. acknowledges the use of the facilities of the CCT-Rosario Computational Center, member of the High Performance Computing National System (SNCAD, MincyT-Argentina).
\end{acknowledgements}

\bibliography{bibliography}

\end{document}